\documentclass[aps,pra,reprint,showpacs]{revtex4-1}

\usepackage{filecontents,amsmath,graphicx,dcolumn,hyperref}
\usepackage{hyperref}
\hypersetup{
    colorlinks=true,       
    linkcolor=blue,          
    citecolor=blue,        
    urlcolor=blue           
}

\begin{document}
\frenchspacing

\title{van der Waals coefficients for positronium interactions with atoms}
\author{A. R. Swann}
\email[Email: ]{aswann02@qub.ac.uk}
\author{J. A. Ludlow}
\altaffiliation[Present address: ]{AquaQ Analytics, Suite 5, Sturgeon Building,
9--15 Queen Street, Belfast BT7 1NN, United Kingdom}
\author{G. F. Gribakin}\email[Email: ]{g.gribakin@qub.ac.uk}
\affiliation{Centre for Theoretical Atomic, Molecular, and Optical Physics,
School of Mathematics and Physics, Queen's University Belfast, Belfast BT7 1NN,
United Kingdom}

\begin{abstract}
The random-phase approximation with exchange (RPAE) is used with a $B$-spline
basis to compute dynamic dipole polarizabilities of noble-gas atoms and several
other closed-shell atoms (Be, Mg, Ca, Zn, Sr, Cd, and Ba). From these, values of
the van der Waals $C_6$ constants for positronium interactions with these atoms
are determined and compared with existing data. Our best predictions of $C_6$
for Ps--noble-gas pairs are expected to be accurate to within 1\%, and to
within a few per cent for the alkaline earths. We also used accurate dynamic
dipole polarizabilities from the literature to compute the $C_6$ coefficients
for the alkali-metal atoms. Implications of increased $C_6$ values for Ps
scattering from more polarizable atoms are discussed.
\end{abstract}


\maketitle

\section{Introduction}

The interaction of positronium (Ps) with matter and antimatter is an important
topic \cite{Ps_review} with applications in many areas of physics. For instance,
the proposed AEgIS experiment would produce an antihydrogen beam from the
reaction between Ps and antiprotons \cite{Kellerbauer2008351,C2CS35454C}. The
antihydrogen would then be used to determine whether antimatter is affected by
gravity in the same manner as matter. Ps is widely used in condensed matter
physics to determine pore sizes in microporous materials and probe
intermolecular voids in polymers \cite{Gidley2006}. Further, positronium
formation in porous materials is used to study its interactions with gases,
e.g., xenon \cite{PhysRevA.88.012511,PhysRevA.88.042517}, or the interaction
between the Ps atoms themselves, with prospects of Bose-Einstein condensation
at room temperature \cite{Mills07,CasMil11}.

Here we focus on the problem of Ps-atom scattering. Compared with electron-atom
scattering and positron-atom scattering, Ps-atom scattering is more difficult
to treat theoretically, chiefly because both scattering objects have an
internal structure \cite{0370-1298-67-8-306}. 

In this paper we address low-energy Ps scattering from closed-shell atoms.
The short-range Ps-atom interaction is repulsive, because (a) the Pauli
principle prevents the electron from Ps from entering the volume occupied by
the atomic electrons, and (b) the positron is repelled by the screened
potential of the atomic nucleus. However, low-energy Ps-atom scattering is also
affected by the attractive long-range van der Waals interaction
\cite{PhysRevA.65.012509,PhysRevA.67.034502,PhysRevA.90.052717}.

The van der Waals potential behaves asymptotically as 
\begin{equation}\label{eq:C6}
U(R) \simeq -\frac{C_6}{R^6},
\end{equation}
where $R$ is the distance between the centers of mass of the atoms, and $C_6$
is the van der Waals coefficient for the atomic pair \cite{landau}. The value
of the $C_6$ constant determines the scattering phase shifts for the partial
waves $l \geq 2$ at low collision momenta $k$ \cite{PhysRevA.5.1684},
\begin{equation}
\delta_l(k) \simeq \frac{6 C_6 \pi k^4}{(2l-3)(2l-1)(2l+1)(2l+3)(2l+5)}.
\end{equation}
The magnitude of $C_6$ also affects the Ps-atom scattering length $A$.
This can be seen from the estimate which uses the potential
(\ref{eq:C6}) with a cut-off at $R=R_0$ \cite{PhysRevA.90.052717},
\begin{equation}\label{eq:A}
A=\left(\frac{mC_6}{8}\right)^{1/4}\frac{\Gamma (3/4)}{\Gamma (5/4)}\frac{J_{-1/4}(x_0)}{J_{1/4}(x_0)},
\end{equation}
where $m$ is the reduced mass, $\Gamma $ and $J_\nu$ are the gamma and Bessel
functions, respectively, and
\begin{equation}\label{eq:x0}
x_0 = \frac{\sqrt{mC_6/2}}{R_0^2}.
\end{equation}
The dimensional prefactor in Eq.~(\ref{eq:A}) determines the characteristic
magnitude of the scattering length in atomic collisions \cite{GF93}.
We use atomic units throughout, so that $m=2$ for Ps-atom collisions.

Mitroy and Bromley \cite{PhysRevA.68.035201} calculated the $C_6$ constants for
Ps--noble-gas interactions using the oscillator strength sum rule. For He they
calculated the oscillator strengths via the configuration interaction (CI)
method, while for Ne, Ar, Kr, and Xe recourse was made to a set of published
pseudo-excitation energies and dipole oscillator strengths \cite{KM85},
leading to semi-empirical (SE) values of $C_6$. In a recent paper devoted to Ps
scattering from Ar and Kr \cite{PhysRevA.90.052717}, Fabrikant and Gribakin
used the London formula \cite{London} for the van der Waals coefficients, which
gave values about 6\% greater than those from
Ref.\ \cite{PhysRevA.68.035201}. To estimate the effect of this difference on
Ps-atom scattering we can use Eq.\ \eqref{eq:A}, which shows that for Ar
($R_0=2.67$ a.u. \cite{PhysRevA.90.052717}), changing $C_6$ from 152 a.u.\
\cite{PhysRevA.90.052717} to 144.1 a.u.\ \cite{PhysRevA.68.035201} increases
the scattering length by about 4\%.
This example shows that Ps-atom scattering is sensitive to the value of the
van der Waals coefficient.

In this work, we employ the random-phase approximation with exchange (RPAE)
\cite{amusiaJETP2,atomicphotoeffect} to calculate the dynamic dipole
polarizabilities of the noble-gas atoms and several other closed-shell atoms
with an $ns^2$ valence shell. These are used to calculate the $C_6$ constants
for Ps-atom pairs \emph{ab initio}. The RPAE method, often called simply the
random-phase approximation (RPA), is equivalent to the (linearized)
time-dependent coupled Hartree-Fock method (see, e.g., Ref.~\cite{TDHF} and
references therein). RPAE is known to provide a good
description of atomic polarization for the noble gases
\cite{amusiaJETP,LudlowPhD}. To make our predictions more accurate, we use
scaling based on the known values of the static dipole polarzabilities.
As a result, we expect our final $C_6$ constants to be accurate to better than
1\%. As far as we are aware, there are currently no published values of $C_6$
for Ps interactions with other closed-shell atoms. Although the RPAE method
is less accurate for them, our final $C_6$ values provide a good
benchmark for future work.

\section{Theory}

The van der Waals coefficient for a pair of atoms, $A$ and $B$, may be
expressed as
\begin{equation}\label{eq:C6AB}
C_6=\frac{2}{3}\sum_{n,n'}\frac{|\langle n_A|{\bf D}|0_A\rangle |^2
|\langle n'_B|{\bf D}|0_B\rangle |^2} {E^A_n+E^B_{n'}-E_0^A-E_0^B}
\end{equation}
where ${\bf D}$ is the electric dipole operator, the matrix elements are taken
between the ground ($0$) and excited ($n$ or $n'$) states of the atoms, with
respective energies $E_0$ and $E_n$, and the extra indices ($A$ or $B$) are used
to distinguish the two atoms \cite{LandauQED}. 

There exists a useful relationship connecting dynamic polarizabilities of
imaginary frequencies with $C_6$. The dynamic polarizability at frequency
$\omega$ is given by
\begin{equation}\label{eqn:dynpol}
\alpha(\omega ) = \frac{1}{3}\sum_{n}\left( \frac{|\langle n|{\bf D}|0\rangle |^2 } {E_n-E_0 - \omega -i\delta }
+\frac{|\langle n|{\bf D}|0\rangle |^2 } {E_n-E_0 +\omega +i\delta }\right),
\end{equation}
where $\delta$ is a positive infinitesimal, which determines the sign of the
imaginary part of $\alpha(\omega)$ for real frequencies $\omega$ above the
ionization potential of the atom. The ground-state static polarizability is
$\alpha(0)$. The van der Waals constant can be obtained by integrating the
product of dynamic polarizabilities of atoms $A$ and $B$ over imaginary
frequencies, viz.
\begin{equation}\label{eqn:c6int}
C_6 = \frac{3}{\pi} \int_0^\infty \alpha_A(i\omega) \alpha_B(i\omega) \, d\omega.
\end{equation}
This result is very convenient because the integration path avoids the poles
of the dynamic polarizabilities.

For hydrogenic atoms (such as H and Ps), the polarizabilities are given by
Eq.\ \eqref{eqn:dynpol} with the dipole matrix elements replaced by the
single-particle radial matrix elements $\langle np | r | 1s \rangle$, where
$r$ is either the electron-proton (H) or electron-positron (Ps) separation.
Thus, for an interacting pair of
hydrogenic atoms, an essentially exact value of $C_6$ can be found. However,
for many-electron atoms (such as the noble gases) the single-particle (e.g.,
Hartree-Fock) method does not give accurate results. RPAE is a many-body theory
method that is known to give accurate dipole polarizabilities and
photoionization cross sections for closed-shell atoms, with the best results
for the noble gases \cite{amusiaJETP,amusiaJETP2}.

After the RPAE equations have been solved (see Appendix \ref{sec:RPAEequations}
for details), the dynamic dipole polarizability is calculated as
\begin{align}\label{eqn:rpaepolarizability}
\alpha(\omega) = - \frac{2}{3} &\left( \sum_{\nu>F,\mu \leq F}
\frac{\langle \mu\|d\|\nu \rangle \langle \nu \|A(\omega)\|\mu \rangle }
{\omega - \epsilon_\nu + \epsilon_\mu }\right. \nonumber \\
&\left. +\sum_{\nu\leq F,\mu > F} \frac{\langle \mu \|d\|\nu \rangle
\langle \nu \|A(\omega)\|\mu \rangle }
{-\omega + \epsilon_\nu - \epsilon_\mu } \right),
\end{align}
where  $\langle \nu \| d \| \mu \rangle$ is the reduced Hartree-Fock dipole
matrix element, $\langle \nu \| A(\omega) \| \mu \rangle$ is the reduced
RPAE dipole matrix element, $\epsilon_{\nu}$ is the energy of state $\nu$, and
$i\delta $ has been dropped for $\omega =0$ and imaginary frequencies. In
Eq.\ \eqref{eqn:rpaepolarizability} the sums over the magnetic quantum numbers
and spins have already been carried out, and the remaining sums are over the
occupied ($\leq F$) or empty ($>F$) electron orbitals $\nu $ and $\mu$.

Note that the matrix element $\langle \nu \| A(\omega) \| \mu \rangle$ is
calculated off-mass-shell, i.e., for $\omega \neq \epsilon_\nu - \epsilon_\mu $,
and
the energy differences in the denominator of Eq.\ \eqref{eqn:rpaepolarizability}
contain the unperturbed Hartree-Fock energies. An equivalent but more
complicated approach would be to calculate these amplitudes on the mass-shell,
simultaneously with finding the RPAE excitation energies $\omega _{\nu \mu }$.
In this case the expression for the polarizability would contain modulus
squared values of $\langle \nu \| A(\omega _{\nu \mu }) \| \mu \rangle$ and the
RPAE excitation energies in the denominator [cf. Eq.\ \eqref{eqn:dynpol}].

The expressions for $C_6$ and $\alpha(\omega)$ contain sums over the complete
sets of excited states. In real systems, such as atoms, these sets of excited
states include both discrete, Rydberg states and the continuum of states with
energies above the ionization potential of the system. By using $B$-splines in
a box of finite radius $R$ (see Section \ref{sec:numericalresults}), the
continuum is discretized in a way that allows accurate numerical calculations
of both the static polarizability (for $\omega=0$) and
$\alpha(i\omega)$, and $C_6$.

\section{\label{sec:numericalresults}Numerical results}

We use $B$-splines to construct either hydrogenic or Hartree-Fock basis states.
They provide an accurate representation of the ground-state orbitals and an
effective spanning of the continuum, due to an appropriately chosen radial
grid (see, e.g., Ref.\ \cite{PhysRevA.70.032720}). In this work we use a set of
60 $B$-splines of order 9, with a box size of $R=15$ a.u. The absence of true
continuum states does not affect the accuracy of $\alpha(0)$ or
$\alpha(i\omega)$ because all excitations in the sums are virtual (the
denominator never vanishes); such virtually excited electrons cannot travel
away from the atom, so describing them by a set of states in a box is accurate.

We calculated static  dipole polarizabilities of H, Ps,  the noble gases, and
other closed-shell atoms and compared with exact theoretical or best
experimental or theoretical values to verify the validity of the method. Then
we computed dynamic polarizabilties $\alpha(i\omega_j)$ over a discrete set of
imaginary frequencies
\begin{equation}
\omega_j = \omega_0 [e^{\sigma(j-1)}-1], \qquad j=1,\dots,N_\omega,
\end{equation} 
where $\omega_0$, $N_\omega$, and $\omega_\text{max} \equiv \omega_{N_\omega}$ are
parameters, and
\begin{equation}
\sigma = \frac{1}{N_\omega - 1} \ln \left( \frac{\omega_\text{max}}
{\omega_0} + 1 \right).
\end{equation}
Values of $\omega_0=0.01$ a.u., $N_\omega=100$, and $\omega_\text{max}=1000$~a.u.\ 
have been used throughout; these values were chosen to provide accurate,
converged values of the integral (\ref{eqn:c6int}). For H and Ps the
polarizabilities were calculated essentially exactly; for the noble gases and
$ns^2$ atoms they were calculated using the RPAE method. The values of $C_6$
were then found by evaluating Eq.\ (\ref{eqn:c6int}) numerically.

The static polarizability obtained for H was $\alpha(0)=4.500$~a.u., in perfect
agreement with the exact value of $\alpha(0)=9/2$ \cite{LandauQED}. The
$B$-spline states and dipole amplitudes for hydrogen can be used to calculate
$\alpha(0)$ and $\alpha(i\omega)$ for Ps by halving the energies and doubling
the amplitudes (due to the reduced mass of Ps being a half of that of H). This
gives $\alpha(0)=36$~a.u.\ for Ps. 

Table \ref{tab:polar} shows the static polarizabilities
obtained using RPAE for the noble-gas and other closed-shell atoms.
\begin{table}
\caption{\label{tab:polar}Static dipole polarizabilities $\alpha(0)$ for the 
noble gases and other closed-shell atoms (in atomic units).}
\begin{ruledtabular}
\begin{tabular}{lddddd}
Atom & \multicolumn{1}{c}{Present} & \multicolumn{1}{c}{RPAE\footnote{This implementation
\cite{amusiaJETP,csap} suffers from poor convergence.}} 
& \multicolumn{1}{c}{RRPA\footnote{Relativistic random-phase approximation \cite{PhysRevA.26.19}.}} 
& \multicolumn{1}{c}{Rec.\footnote{Recommended experimental values for the noble
gases, Zn, and Cd \cite{onlinepolarizabilities}, and
calculated values for alkaline earth metals \cite{porsevJETP}.}} 
& \multicolumn{1}{c}{$F$\footnote{$F$ is the ratio of the recommended value
to the present value.}} \\
\hline
He & 1.322 & 1.256 & 1.322 & 1.384 & 1.047 \\
Ne & 2.377 & 2.30 & 2.38 & 2.67 & 1.123 \\
Ar & 10.758 & 10.73 & 10.77 &11.07 & 1.029 \\
Kr & 16.476 & 16.18 & 16.47 &17.075 & 1.036 \\
Xe & 27.099 & 27.98 & 26.97 & 27.815 & 1.026 \\
\hline
Be & 45.604 & 43.2 & 45.6 & 37.76 & 0.828 \\
Mg & 81.502 & \text{---} & 81.2 & 71.3 & 0.875 \\
Ca & 183.965 & 166 & 182.8 & 157.1 & 0.854 \\
Zn & 54.046 & \text{---}& 50.8 & 38.8 & 0.718 \\
Sr & 242.240 & \text{---}& 232.6 & 197.2 & 0.814 \\
Cd & 75.958 & \text{---}& 63.7 & 49.65 & 0.654 \\
Ba & 355.735 & \text{---}& 324.0 & 273.5 & 0.769 \\
\end{tabular}
\end{ruledtabular}
\end{table}
The RPAE static polarizabilities for the noble-gas atoms agree to 0.1\% or
better with the results of the equivalent coupled Hartree-Fock calculation
\cite{McESG79}, with the present values for Kr and Xe being more accurate
numerically due to better convergence. The results are also in agreement with
calculations made using the relativistic
random-phase approximation (RRPA, which also accounts for exchange). Larger
differences are to be expected for heavier atoms, since relativistic
corrections scale as $(Z/137)^2$, where $Z$ is the nuclear charge. However, even
for the heaviest noble-gas atom (Xe) the difference is less than 0.5\%. This
bodes well for using the dynamic polarizabilities to calculate the $C_6$
coefficients for Ps--noble-gas pairs. As expected, for the $ns^2$ atoms, the
agreement with RRPA is poorer for heavier atoms (since the $s$ electrons are
affected more strongly by the relativistic corrections). The agreement with the
experimental values for these atoms is also poorer in general. This is related
to the smaller ionization potentials and larger effect of the non-RPA
correlation effects (e.g., two-hole-two-particle excitations) in these systems.

The results for $C_6$ are displayed in Table \ref{tab:c6}.
\begin{table}
\caption{\label{tab:c6}van der Waals $C_6$ coefficients for various Ps-$X$
systems (in atomic units).}
\begin{ruledtabular}
\begin{tabular}{ldddd}
System & \multicolumn{1}{c}{Present} & \multicolumn{1}{c}{Best
prediction\footnote{Obtained by scaling the matrix elements by $\xi$ and
energies by $1/\xi^2$, where $\xi=F^{1/4}$ (see Table \ref{tab:polar}).}} 
&\multicolumn{1}{c}{CI/SE\footnote{Computed in Ref.~\cite{PhysRevA.68.035201} using the configuration-interaction (CI) for He and semi-empirical dipole polarizabilities \cite{KM85} for the noble gases, and in the present work, using empirically-adjusted CI + many-body theory data \cite{Derevianko2010323} for the alkaline-earth atoms.}}
& \multicolumn{1}{c}{Other\footnote{For Ps-H and Ps-Ps: pseudostate calculations \cite{0022-3700-13-17-017}.  
For all other systems, these values were calculated using the London formula $C_6 = 3\alpha_A\alpha_B I_A I_B /[2(I_A + I_B)]$, 
where $\alpha_{A,B}$ and $I_{A,B}$ are the static dipole polarizabilities and ionization potentials of the atoms \cite{London}; 
the polarizabilities were taken from Ref.\ \cite{onlinepolarizabilities} and the ionization potentials were taken from Ref.\ \cite{CRC}.}} \\
\hline
Ps-H & 34.785 &\text{---} & 34.785 & 34.785 \\
Ps-Ps & 207.969 &\text{---} & \text{---} & 207.966\\
\hline
Ps-He & 12.849 & 13.41 & 13.37 & 14.6 \\
Ps-Ne & 23.759 & 26.48 & 26.74 & 27.4\\
Ps-Ar & 96.212 & 98.69 & 98.50 & 104.4\\
Ps-Kr & 142.185 & 146.71 & 144.1 & 155.1\\
Ps-Xe & 222.355 & 227.38 & 221.6 & 240.6 \\
\hline
Ps-Be & 241.416 & 208.6 & 210.7 & 294.3 \\
Ps-Mg & 393.484 & 355.8 & 357.9 & 506.4 \\
Ps-Ca & 717.138 & 641.5 & 636.6 & 1079 \\
Ps-Zn & 300.756 & 231.4 & \text{---} & 303.8 \\
Ps-Sr & 885.425 & 767.5 & 763.8 & 1144 \\
Ps-Cd & 397.161 & 285.0 & \text{---} & 381.6 \\
Ps-Ba & 1164.429 & 975.6& 966.8 & 1569 \\
\end{tabular}
\end{ruledtabular}
\end{table}
Our calculations which use RPAE polarizabilities (``Present'' in Table
\ref{tab:c6}) are in close agreement with the SE calculations of Mitroy and
Bromley \cite{PhysRevA.68.035201} for the heavier noble gases; the relative
differences for Ps-Ar, Ps-Kr, and Ps-Xe are 2\%, 1\%, and 0.3\% respectively.
For the lighter noble gases the differences are more significant: 4\% and
11\% for Ps-He and Ps-Ne, respectively. This discrepancy can be traced back to
the fact that the RPAE polarizabilities for He and Ne are lower than the
recommended values, by 5\% and 12\%, respectively. Correcting the RPAE $C_6$
value for He by the corresponding factor $F$ (see Table \ref{tab:polar}) gives
$C_6=13.45$, which agrees with the accurate value for He from
Ref.\ \cite{PhysRevA.68.035201} to within 0.6\%. However, this crude scaling
violates the Thomas-Reiche-Kuhn sum rule, which in the limit of large $\omega$
gives
\begin{equation}\label{eqn:sumrule}
\alpha(i\omega) \simeq \frac{N}{\omega^2},
\end{equation}
where $N$ is the number of atomic electrons. A better approach is to scale the
dipole matrix elements $\langle \mu\|d\|\nu \rangle$ and
$ \langle \nu \|A(\omega)\|\mu \rangle$ in Eq.\ (\ref{eqn:rpaepolarizability})
by some factor $\xi$ and the orbital energies in the denominators by $1/\xi^2$.
In this case the sum rule is preserved, while the calculated static
polarizability is scaled by $\xi^4$. The appropriate choice of $\xi$ is given
by $\xi = F^{1/4}$. Carrying out this scaling for He gives $C_6=13.41$, which
agrees with the value from Ref.\ \cite{PhysRevA.68.035201} to within 0.3\%.
This is a clear improvement over the simple scaling by a factor of $F$.

As a further test of the effectiveness of this scaling, the $C_6$ constants
were calculated for pairs of the noble-gas and  alkaline-earth-metal atoms.
These were compared with data from Ref.\ \cite{Derevianko2010323}, wherein
several relativistic many-body theory \cite{porsevJETP,PhysRevA.65.020701} and
semi-empirical \cite{kumar1985pseudo} methods were employed. For noble-gas
pairs, the minimum relative difference was 0.7\% (for Ar-Ar), while the
maximum (for Xe-Xe) was 5\% (due to the use of the dynamical polarizability
``normalized'' to $\alpha (0)=27.16$~a.u. \cite{kumar1985pseudo} in
Ref.~\cite{Derevianko2010323},
which is lower than the recommended experimental value $\alpha (0)=27.815$~a.u.
that we use). For alkaline-earth-metal pairs, the comparison
was actually better; the relative differences ranged from 0.2\% (for Sr-Sr) to
2.1\% (for Be-Be). This is because we use the same recommended static
polarizabilities for the alkaline earth atoms as in Ref.\ \cite{Derevianko2010323}.

Scaling the \textit{ab initio} RPAE polarizabilities in this way produces our
best prediction of the $C_6$ van der Waals coefficients (third column in
Table \ref{tab:c6}). We expect that for the noble-gas atoms these values are
accurate to within 1\%. The relative differences from the CI/SE
values of Mitroy and Bromley for Ne, Ar, Kr, and Xe are 1\%, 0.2\%, 1.8\%, and
2.6\%, respectively. We believe that for Kr and Xe our best prediction values
are superior to those of Ref.\ \cite{PhysRevA.68.035201}, where the
polarizabilities from Ref.\ \cite{kumar1985pseudo} were used.

Since the RPAE polarizabilities of the $ns^2$ atoms are greater than
the recommended values of $\alpha (0)$ by 15--50\%, the calculated
\textit{ab initio} $C_6$ values significantly overestimate the true van der 
Waals coefficients. Given the larger discrepancy for $\alpha (0)$, the
use of scaling is a cruder procedure for improving the $C_6$
constants. In this case we believe that our best predictions
are accurate within few per cent for the alkaline earth atoms, and within
5--10\% for Zn and Cd.

As a further test of the accuracy of our predictions for the alkaline-earth
atoms, we used tabulated dynamic dipole polarizabilities
from Ref.~\cite{Derevianko2010323}, which were computed using a combination of
relativistic methods, including RRPA, CI and many-body perturbation theory,
and further adjusted using accurate theoretical and experimental data. These
values are presented in the fourth column of Table~\ref{tab:c6} (lower half).
In all cases the difference between these values and our best predictions does
not exceed 1\%. Ref.~\cite{Derevianko2010323} also presents accurate dynamic
dipole polarizabilities for the alkali-metal atoms, and we used these to
compute the $C_6$ coefficients for the Ps-alkali-metal pairs, see
Table~\ref{tab:alk}. Comparison with semi-empirical calculations presented in 
Ref.~\cite{PhysRevA.68.035201} shows excellent agreement.

\begin{table}
\caption{Van der Waals $C_6$ coefficients for Ps-alkali-metal systems (in atomic units).}
\label{tab:alk}
\begin{ruledtabular}
\begin{tabular}{ccccccc}
Atom & Li & Na & K & Rb & Cs & Fr \\
\hline
$C_6$\footnote{Computed in the present work using empirically-adjusted many-body theory data \cite{Derevianko2010323}.} & 462.8 & 505.7 & 765.7 & 846.8 & 1014.6 & 937.5 \\
$C_6$\footnote{Semi-empirical calculations from Ref.~\cite{PhysRevA.68.035201}.} & 463.9 & 507.2 & 768.3 & 844.8 & \text{---}  & \text{---} \\
\end{tabular}
\end{ruledtabular}
\end{table}

Looking at the last column in Table \ref{tab:c6}, we see that the London
formula does a reasonable job for the more rigid noble-gas atoms, but tends
to overestimate the $C_6$ coefficients significantly for Ps interacting with
$ns^2$ atoms.

\section{Conclusions}

Dynamic dipole polarizabilities over a range of imaginary frequencies were
computed numerically exactly for H and Ps, and by using RPAE for the noble-gas
atoms and several other closed-shell atoms. The static polarizabilities for
the noble gases deviated from relativistic (RRPA) calculations by no more than
0.5\%, and for Ar, Kr, and Xe were within few per cent of experimental values.
There was greater error for the other closed-shell atoms, but this was
expected; the RPAE method is most suitable for the noble gases.

\emph{Ab initio} calculations of the van der Waals $C_6$ coefficients for Ps
interactions with these atoms were performed. For the heavier noble gases, close
agreement is observed with previous calculations \cite{PhysRevA.68.035201}.
For the lighter noble gases (He and Ne) the discrepancies are more significant,
which motivates a sum-rule preserving scaling of the dynamical polarizabilities
to calculate $C_6$ values. Though these data are no longer \emph{ab initio},
they are expected to the be the most accurate values currently available. For
the most part, our values of $C_6$ for Ps interactions with the $ns^2$ atoms
differ significantly from the London values. Here our best predicitons are
less accurate (few to 10\%), but as there are no other theoretical calculations
of these data, these values will provide a useful benchmark for future
calculations.

It is interesting to discuss the implications of the $C_6$ values for low-energy
Ps-atom scattering. It is clear from Table~\ref{tab:c6} that for more
polarizable (and more weakly bound) atoms, the $C_6$ values are greater, and
that Ps will experience a stronger van der Waals attraction to these atoms.
However, it can be seen from the parameter $x_0$, Eq.~(\ref{eq:x0}), which
determines the scattering length in Eq.~(\ref{eq:A}), that the increase in
$C_6$ for such atoms is offset by the increase in the parameter $R_0$, which is
proportional to the atomic radius. In fact, the latter effect makes $x_0$
smaller for the more weakly bound atoms. As a result, the more polarizable
atoms are not more attractive for Ps, and will likely have larger positive Ps
scattering lengths,
due to their larger geometric sizes. This also shows that Ps binding to
closed-shell neutral atoms does not occur. Note that this is in contrast to
many open-shell atoms, e.g., Na, Cu, or the halogens, which do bind Ps by
accommodating the extra electron in their valence shell \cite{MBR02,halogen}.
However, even in the case of Ps bound states with alkali-metal atoms, the 
binding energy decreases in the sequence PsLi, PsNa, PsK \cite{MR99}, in spite
of the greater values of the $C_6$ constant. This is primarily the effect of
the increasing atomic radius.

It is hoped that the results presented here will be useful for studies of
Ps-noble-gas-atom scattering and Ps interactions with $ns^2$ and alkali-metal
atoms. With little extra work, the method can be extended to calculate
quadrupole and higher polarizabilities and determine higher-order van der
Waals coefficients $C_8$ and $C_{10}$.

\begin{acknowledgments}
The work of A.R.S. has been supported by the Department of Employment and
Learning (Northern Ireland). 
\end{acknowledgments}

\appendix

\section{RPAE equations}\label{sec:RPAEequations}

In the equations for the RPAE dipole matrix element
$\langle \nu | A(\omega) | \mu \rangle$, we must distinguish between the hole
states, i.e., states below the Fermi level $F$, and particle states, i.e.,
states above the Fermi level $F$. For $\nu>F$ and $\mu \leq F$ we have
\begin{widetext}
\begin{align}\label{eqn:rpae1}
\langle \nu | A(\omega) | \mu \rangle = \langle \nu | d | \mu \rangle
+ \left( \sum_{\nu'>F,\mu'\leq F} -
\sum_{\mu'>F,\nu'\leq F} \right)
\frac{\langle \nu \mu'|V|\nu'\mu\rangle -\langle \mu' \nu |V|\nu' \mu \rangle}
{\omega - \epsilon_{\nu'} + \epsilon_{\mu'} + i(1-2n_{\nu'})\delta}
\langle \nu' | A(\omega) | \mu' \rangle,
\end{align}
and a formally identical equation for $\nu \leq F$ and $\mu > F$.

In Eq.\ (\ref{eqn:rpae1}), $\langle \nu | d | \mu \rangle$ is the Hartree-Fock
dipole matrix element; $\langle \nu\mu'|V|\nu'\mu\rangle$ is the Coulomb matrix
element, which is defined by
\begin{equation}
\langle \nu\mu'|V|\nu'\mu\rangle \equiv \iint \varphi_\nu^*({\bf r}) \varphi_{\mu'}^*({\bf r'}) \frac{1}{\vert {\bf r} - {\bf r'} \vert} \varphi_{\nu'}({\bf r'}) \varphi_{\mu}({\bf r})\,d^3 {\bf r}\, d^3 {\bf r'},
\end{equation}
where the $\varphi_\nu $ are single-particle wave functions; $\epsilon_{\nu'}$ is
the energy of state $\nu'$; and
\begin{equation}
n_{\nu'} = \begin{cases}
0 & \text{for}\quad \nu'>F, \\
1 & \text{for} \quad \nu' \leq F.
\end{cases}
\end{equation}

By separating the angular and radial parts in the electronic states
$\varphi $, and then integrating over the angular variables and summing over
the magnetic quantum numbers and spins, one obtains the RPAE equations for
the \textit{reduced} amplitudes $\langle \nu \| A(\omega) \| \mu \rangle $
in the form
\begin{equation}\label{eqn:rpaereduced}
\langle \nu \| A(\omega) \| \mu \rangle = \langle \nu \| d \| \mu \rangle
+
\frac{1}{3} \left( \sum_{\nu'>F,\mu'\leq F} - \sum_{\mu'>F,\nu'\leq F} \right)
\frac{\langle \nu \mu' \| U_1 \| \nu' \mu \rangle \langle \nu' \| A(\omega) \| \mu' \rangle}
{\omega - \epsilon_{\nu'} + \epsilon_{\mu'} + i(1-2n_{\nu'})\delta},
\end{equation}
where
\begin{equation}
\langle \nu \| d \| \mu \rangle = (-1)^{l_\nu} \sqrt{[l_\nu] [l_\mu]}
\begin{pmatrix} l_\nu & 1 & l_\mu \\ 0 & 0 & 0 \end{pmatrix}
\int_0^R P_\nu(r) r P_\mu(r) \, dr,
\end{equation}
\begin{equation}\label{eqn:U1}
\langle \nu \mu' \| U_1 \| \nu' \mu \rangle = 2 \langle \nu \mu' \| V_1 \| \nu' \mu \rangle
- 3 \sum_{l=0}^\infty
(-1)^{l-1}
\begin{Bmatrix} l_\nu & 1 & l_\mu \\ l_{\mu'} & l & l_{\nu'} \end{Bmatrix} \langle \nu \mu' \| V_l \| \mu \nu' \rangle ,
\end{equation}
and
\begin{equation}
\langle \nu\mu'\| V_l \|\nu'\mu\rangle = \sqrt{[l_\nu] [l_{\mu'}] [l_{\nu'}] [l_\mu]} 
\begin{pmatrix} l_\nu & l & l_\mu \\ 0 & 0 & 0 \end{pmatrix}
\begin{pmatrix} l_{\mu'} & l & l_{\nu'} \\ 0 & 0 & 0 \end{pmatrix}
\int_0^R \int_0^R P_\nu(r) P_{\mu'}(r') \frac{r_<^l}{r_>^{l+1}} P_{\nu'}(r') P_\mu(r) \, dr \, dr',
\end{equation}
are the reduced dipole and Coulomb matrix elements, $P_\nu (r)$ are radial wave
functions, $[l_\nu]\equiv 2l_\nu+1$, $r_>=\max(r,r')$, $r_<=\min(r,r')$, and $R$
is the box radius in our $B$-spline-basis implementation.
\end{widetext}

Introducing vectors $x$ and $y$ for $\langle \nu \| A(\omega) \| \mu \rangle$ for $\nu>F$, $\mu\leq F$ and $\nu\leq F$, $\mu > F$, respectively, we 
can write equations (\ref{eqn:rpaereduced}) in block matrix form as 
\begin{equation}\label{eqn:rpaematrix}
\begin{pmatrix} x \\ y \end{pmatrix} = \begin{pmatrix} d \\ d \end{pmatrix} + \begin{pmatrix} U_{1a} & U_{1b} \\ U_{1b} & U_{1a} \end{pmatrix} \begin{pmatrix} \chi_1 & 0 \\ 0 & \chi_2 \end{pmatrix} \begin{pmatrix} x \\ y \end{pmatrix},
\end{equation}
where $\chi_1$ and $\chi_2$ are the diagonal matrices of energy denominators,
$d$ is the vector of Hartree-Fock dipole matrix elements, and the matrices
$U_{1a}$ and $U_{1b}$ represent the two terms in \eqref{eqn:U1}. This linear
equation can be solved numerically for the RPAE dipole amplitudes, and then the
dynamic dipole polarizability is calculated from Eq.\ \eqref{eqn:rpaepolarizability}.

\bibliography{c6_biblio.bib}

\begin{thebibliography}{38}%
\makeatletter
\providecommand \@ifxundefined [1]{%
 \@ifx{#1\undefined}
}%
\providecommand \@ifnum [1]{%
 \ifnum #1\expandafter \@firstoftwo
 \else \expandafter \@secondoftwo
 \fi
}%
\providecommand \@ifx [1]{%
 \ifx #1\expandafter \@firstoftwo
 \else \expandafter \@secondoftwo
 \fi
}%
\providecommand \natexlab [1]{#1}%
\providecommand \enquote  [1]{``#1''}%
\providecommand \bibnamefont  [1]{#1}%
\providecommand \bibfnamefont [1]{#1}%
\providecommand \citenamefont [1]{#1}%
\providecommand \href@noop [0]{\@secondoftwo}%
\providecommand \href [0]{\begingroup \@sanitize@url \@href}%
\providecommand \@href[1]{\@@startlink{#1}\@@href}%
\providecommand \@@href[1]{\endgroup#1\@@endlink}%
\providecommand \@sanitize@url [0]{\catcode `\\12\catcode `\$12\catcode
  `\&12\catcode `\#12\catcode `\^12\catcode `\_12\catcode `\%12\relax}%
\providecommand \@@startlink[1]{}%
\providecommand \@@endlink[0]{}%
\providecommand \url  [0]{\begingroup\@sanitize@url \@url }%
\providecommand \@url [1]{\endgroup\@href {#1}{\urlprefix }}%
\providecommand \urlprefix  [0]{URL }%
\providecommand \Eprint [0]{\href }%
\providecommand \doibase [0]{http://dx.doi.org/}%
\providecommand \selectlanguage [0]{\@gobble}%
\providecommand \bibinfo  [0]{\@secondoftwo}%
\providecommand \bibfield  [0]{\@secondoftwo}%
\providecommand \translation [1]{[#1]}%
\providecommand \BibitemOpen [0]{}%
\providecommand \bibitemStop [0]{}%
\providecommand \bibitemNoStop [0]{.\EOS\space}%
\providecommand \EOS [0]{\spacefactor3000\relax}%
\providecommand \BibitemShut  [1]{\csname bibitem#1\endcsname}%
\let\auto@bib@innerbib\@empty
\bibitem [{\citenamefont {Laricchia}\ and\ \citenamefont
  {Walters}(2012)}]{Ps_review}%
  \BibitemOpen
  \bibfield  {author} {\bibinfo {author} {\bibfnamefont {G.}~\bibnamefont
  {Laricchia}}\ and\ \bibinfo {author} {\bibfnamefont {H.~R.~J.}\ \bibnamefont
  {Walters}},\ }\href {\doibase 10.1393/ncr/i2012-10077-6} {\bibfield
  {journal} {\bibinfo  {journal} {Rivista del Nuovo Cimento}\ }\textbf
  {\bibinfo {volume} {35}},\ \bibinfo {pages} {305} (\bibinfo {year}
  {2012})}\BibitemShut {NoStop}%
\bibitem [{\citenamefont {Kellerbauer}\ \emph {et~al.}(2008)\citenamefont
  {Kellerbauer}, \citenamefont {Amoretti}, \citenamefont {Belov}, \citenamefont
  {Bonomi}, \citenamefont {Boscolo}, \citenamefont {Brusa}, \citenamefont
  {B\"uchner}, \citenamefont {Byakov}, \citenamefont {Cabaret}, \citenamefont
  {Canali}, \citenamefont {Carraro}, \citenamefont {Castelli}, \citenamefont
  {Cialdi}, \citenamefont {de~Combarieu}, \citenamefont {Comparat},
  \citenamefont {Consolati}, \citenamefont {Djourelov}, \citenamefont {Doser},
  \citenamefont {Drobychev}, \citenamefont {Dupasquier}, \citenamefont
  {Ferrari}, \citenamefont {Forget}, \citenamefont {Formaro}, \citenamefont
  {Gervasini}, \citenamefont {Giammarchi}, \citenamefont {Gninenko},
  \citenamefont {Gribakin}, \citenamefont {Hogan}, \citenamefont {Jacquey},
  \citenamefont {Lagomarsino}, \citenamefont {Manuzio}, \citenamefont
  {Mariazzi}, \citenamefont {Matveev}, \citenamefont {Meier}, \citenamefont
  {Merkt}, \citenamefont {Nedelec}, \citenamefont {Oberthaler}, \citenamefont
  {Pari}, \citenamefont {Prevedelli}, \citenamefont {Quasso}, \citenamefont
  {Rotondi}, \citenamefont {Sillou}, \citenamefont {Stepanov}, \citenamefont
  {Stroke}, \citenamefont {Testera}, \citenamefont {Tino}, \citenamefont
  {Tr\'enec}, \citenamefont {Vairo}, \citenamefont {Vigu\'e}, \citenamefont
  {Walters}, \citenamefont {Warring}, \citenamefont {Zavatarelli},\ and\
  \citenamefont {Zvezhinskij}}]{Kellerbauer2008351}%
  \BibitemOpen
  \bibfield  {author} {\bibinfo {author} {\bibfnamefont {A.}~\bibnamefont
  {Kellerbauer}}, \bibinfo {author} {\bibfnamefont {M.}~\bibnamefont
  {Amoretti}}, \bibinfo {author} {\bibfnamefont {A.}~\bibnamefont {Belov}},
  \bibinfo {author} {\bibfnamefont {G.}~\bibnamefont {Bonomi}}, \bibinfo
  {author} {\bibfnamefont {I.}~\bibnamefont {Boscolo}}, \bibinfo {author}
  {\bibfnamefont {R.}~\bibnamefont {Brusa}}, \bibinfo {author} {\bibfnamefont
  {M.}~\bibnamefont {B\"uchner}}, \bibinfo {author} {\bibfnamefont
  {V.}~\bibnamefont {Byakov}}, \bibinfo {author} {\bibfnamefont
  {L.}~\bibnamefont {Cabaret}}, \bibinfo {author} {\bibfnamefont
  {C.}~\bibnamefont {Canali}}, \bibinfo {author} {\bibfnamefont
  {C.}~\bibnamefont {Carraro}}, \bibinfo {author} {\bibfnamefont
  {F.}~\bibnamefont {Castelli}}, \bibinfo {author} {\bibfnamefont
  {S.}~\bibnamefont {Cialdi}}, \bibinfo {author} {\bibfnamefont
  {M.}~\bibnamefont {de~Combarieu}}, \bibinfo {author} {\bibfnamefont
  {D.}~\bibnamefont {Comparat}}, \bibinfo {author} {\bibfnamefont
  {G.}~\bibnamefont {Consolati}}, \bibinfo {author} {\bibfnamefont
  {N.}~\bibnamefont {Djourelov}}, \bibinfo {author} {\bibfnamefont
  {M.}~\bibnamefont {Doser}}, \bibinfo {author} {\bibfnamefont
  {G.}~\bibnamefont {Drobychev}}, \bibinfo {author} {\bibfnamefont
  {A.}~\bibnamefont {Dupasquier}}, \bibinfo {author} {\bibfnamefont
  {G.}~\bibnamefont {Ferrari}}, \bibinfo {author} {\bibfnamefont
  {P.}~\bibnamefont {Forget}}, \bibinfo {author} {\bibfnamefont
  {L.}~\bibnamefont {Formaro}}, \bibinfo {author} {\bibfnamefont
  {A.}~\bibnamefont {Gervasini}}, \bibinfo {author} {\bibfnamefont
  {M.}~\bibnamefont {Giammarchi}}, \bibinfo {author} {\bibfnamefont
  {S.}~\bibnamefont {Gninenko}}, \bibinfo {author} {\bibfnamefont
  {G.}~\bibnamefont {Gribakin}}, \bibinfo {author} {\bibfnamefont
  {S.}~\bibnamefont {Hogan}}, \bibinfo {author} {\bibfnamefont
  {M.}~\bibnamefont {Jacquey}}, \bibinfo {author} {\bibfnamefont
  {V.}~\bibnamefont {Lagomarsino}}, \bibinfo {author} {\bibfnamefont
  {G.}~\bibnamefont {Manuzio}}, \bibinfo {author} {\bibfnamefont
  {S.}~\bibnamefont {Mariazzi}}, \bibinfo {author} {\bibfnamefont
  {V.}~\bibnamefont {Matveev}}, \bibinfo {author} {\bibfnamefont
  {J.}~\bibnamefont {Meier}}, \bibinfo {author} {\bibfnamefont
  {F.}~\bibnamefont {Merkt}}, \bibinfo {author} {\bibfnamefont
  {P.}~\bibnamefont {Nedelec}}, \bibinfo {author} {\bibfnamefont
  {M.}~\bibnamefont {Oberthaler}}, \bibinfo {author} {\bibfnamefont
  {P.}~\bibnamefont {Pari}}, \bibinfo {author} {\bibfnamefont {M.}~\bibnamefont
  {Prevedelli}}, \bibinfo {author} {\bibfnamefont {F.}~\bibnamefont {Quasso}},
  \bibinfo {author} {\bibfnamefont {A.}~\bibnamefont {Rotondi}}, \bibinfo
  {author} {\bibfnamefont {D.}~\bibnamefont {Sillou}}, \bibinfo {author}
  {\bibfnamefont {S.}~\bibnamefont {Stepanov}}, \bibinfo {author}
  {\bibfnamefont {H.}~\bibnamefont {Stroke}}, \bibinfo {author} {\bibfnamefont
  {G.}~\bibnamefont {Testera}}, \bibinfo {author} {\bibfnamefont
  {G.}~\bibnamefont {Tino}}, \bibinfo {author} {\bibfnamefont {G.}~\bibnamefont
  {Tr\'enec}}, \bibinfo {author} {\bibfnamefont {A.}~\bibnamefont {Vairo}},
  \bibinfo {author} {\bibfnamefont {J.}~\bibnamefont {Vigu\'e}}, \bibinfo
  {author} {\bibfnamefont {H.}~\bibnamefont {Walters}}, \bibinfo {author}
  {\bibfnamefont {U.}~\bibnamefont {Warring}}, \bibinfo {author} {\bibfnamefont
  {S.}~\bibnamefont {Zavatarelli}}, \ and\ \bibinfo {author} {\bibfnamefont
  {D.}~\bibnamefont {Zvezhinskij}},\ }\href {\doibase
  10.1016/j.nimb.2007.12.010} {\bibfield  {journal} {\bibinfo  {journal} {Nucl.
  Instrum. Methods B}\ }\textbf {\bibinfo {volume} {266}},\ \bibinfo {pages}
  {351} (\bibinfo {year} {2008})}\BibitemShut {NoStop}%
\bibitem [{\citenamefont {Consolati}\ \emph {et~al.}(2013)\citenamefont
  {Consolati}, \citenamefont {Ferragut}, \citenamefont {Galarneau},
  \citenamefont {Di~Renzo},\ and\ \citenamefont {Quasso}}]{C2CS35454C}%
  \BibitemOpen
  \bibfield  {author} {\bibinfo {author} {\bibfnamefont {G.}~\bibnamefont
  {Consolati}}, \bibinfo {author} {\bibfnamefont {R.}~\bibnamefont {Ferragut}},
  \bibinfo {author} {\bibfnamefont {A.}~\bibnamefont {Galarneau}}, \bibinfo
  {author} {\bibfnamefont {F.}~\bibnamefont {Di~Renzo}}, \ and\ \bibinfo
  {author} {\bibfnamefont {F.}~\bibnamefont {Quasso}},\ }\href {\doibase
  10.1039/C2CS35454C} {\bibfield  {journal} {\bibinfo  {journal} {Chem. Soc.
  Rev.}\ }\textbf {\bibinfo {volume} {42}},\ \bibinfo {pages} {3821} (\bibinfo
  {year} {2013})}\BibitemShut {NoStop}%
\bibitem [{\citenamefont {Gidley}\ \emph {et~al.}(2006)\citenamefont {Gidley},
  \citenamefont {Peng},\ and\ \citenamefont {Vallery}}]{Gidley2006}%
  \BibitemOpen
  \bibfield  {author} {\bibinfo {author} {\bibfnamefont {D.~W.}\ \bibnamefont
  {Gidley}}, \bibinfo {author} {\bibfnamefont {H.-G.}\ \bibnamefont {Peng}}, \
  and\ \bibinfo {author} {\bibfnamefont {R.~S.}\ \bibnamefont {Vallery}},\
  }\href {\doibase 10.1146/annurev.matsci.36.111904.135144} {\bibfield
  {journal} {\bibinfo  {journal} {Ann. Rev. Mat. Res.}\ }\textbf {\bibinfo
  {volume} {36}},\ \bibinfo {pages} {49} (\bibinfo {year} {2006})}\BibitemShut
  {NoStop}%
\bibitem [{\citenamefont {Shibuya}\ \emph
  {et~al.}(2013{\natexlab{a}})\citenamefont {Shibuya}, \citenamefont
  {Nakayama}, \citenamefont {Saito},\ and\ \citenamefont
  {Hyodo}}]{PhysRevA.88.012511}%
  \BibitemOpen
  \bibfield  {author} {\bibinfo {author} {\bibfnamefont {K.}~\bibnamefont
  {Shibuya}}, \bibinfo {author} {\bibfnamefont {T.}~\bibnamefont {Nakayama}},
  \bibinfo {author} {\bibfnamefont {H.}~\bibnamefont {Saito}}, \ and\ \bibinfo
  {author} {\bibfnamefont {T.}~\bibnamefont {Hyodo}},\ }\href {\doibase
  10.1103/PhysRevA.88.012511} {\bibfield  {journal} {\bibinfo  {journal} {Phys.
  Rev. A}\ }\textbf {\bibinfo {volume} {88}},\ \bibinfo {pages} {012511}
  (\bibinfo {year} {2013}{\natexlab{a}})}\BibitemShut {NoStop}%
\bibitem [{\citenamefont {Shibuya}\ \emph
  {et~al.}(2013{\natexlab{b}})\citenamefont {Shibuya}, \citenamefont
  {Kawamura},\ and\ \citenamefont {Saito}}]{PhysRevA.88.042517}%
  \BibitemOpen
  \bibfield  {author} {\bibinfo {author} {\bibfnamefont {K.}~\bibnamefont
  {Shibuya}}, \bibinfo {author} {\bibfnamefont {Y.}~\bibnamefont {Kawamura}}, \
  and\ \bibinfo {author} {\bibfnamefont {H.}~\bibnamefont {Saito}},\ }\href
  {\doibase 10.1103/PhysRevA.88.042517} {\bibfield  {journal} {\bibinfo
  {journal} {Phys. Rev. A}\ }\textbf {\bibinfo {volume} {88}},\ \bibinfo
  {pages} {042517} (\bibinfo {year} {2013}{\natexlab{b}})}\BibitemShut
  {NoStop}%
\bibitem [{\citenamefont {Mills}(2007)}]{Mills07}%
  \BibitemOpen
  \bibfield  {author} {\bibinfo {author} {\bibfnamefont {A.~P.}\ \bibnamefont
  {Mills}, \bibfnamefont {Jr.}},\ }\href {\doibase
  10.1016/j.radphyschem.2006.03.009} {\bibfield  {journal} {\bibinfo  {journal}
  {Rad. Phys. Chem.}\ }\textbf {\bibinfo {volume} {76}},\ \bibinfo {pages} {76}
  (\bibinfo {year} {2007})}\BibitemShut {NoStop}%
\bibitem [{\citenamefont {Cassidy}\ and\ \citenamefont
  {Mills}(2011)}]{CasMil11}%
  \BibitemOpen
  \bibfield  {author} {\bibinfo {author} {\bibfnamefont {D.~B.}\ \bibnamefont
  {Cassidy}}\ and\ \bibinfo {author} {\bibfnamefont {A.~P.}\ \bibnamefont
  {Mills}, \bibfnamefont {Jr.}},\ }\href {\doibase
  10.1103/PhysRevLett.100.013401} {\bibfield  {journal} {\bibinfo  {journal}
  {Phys. Rev. Lett.}\ }\textbf {\bibinfo {volume} {100}},\ \bibinfo {pages}
  {013401} (\bibinfo {year} {2011})}\BibitemShut {NoStop}%
\bibitem [{\citenamefont {Massey}\ and\ \citenamefont
  {Mohr}(1954)}]{0370-1298-67-8-306}%
  \BibitemOpen
  \bibfield  {author} {\bibinfo {author} {\bibfnamefont {H.~S.~W.}\
  \bibnamefont {Massey}}\ and\ \bibinfo {author} {\bibfnamefont {C.~B.~O.}\
  \bibnamefont {Mohr}},\ }\href {http://stacks.iop.org/0370-1298/67/i=8/a=306}
  {\bibfield  {journal} {\bibinfo  {journal} {Proc. Phys. Soc. Section A}\
  }\textbf {\bibinfo {volume} {67}},\ \bibinfo {pages} {695} (\bibinfo {year}
  {1954})}\BibitemShut {NoStop}%
\bibitem [{\citenamefont {Mitroy}\ and\ \citenamefont
  {Ivanov}(2001)}]{PhysRevA.65.012509}%
  \BibitemOpen
  \bibfield  {author} {\bibinfo {author} {\bibfnamefont {J.}~\bibnamefont
  {Mitroy}}\ and\ \bibinfo {author} {\bibfnamefont {I.~A.}\ \bibnamefont
  {Ivanov}},\ }\href {\doibase 10.1103/PhysRevA.65.012509} {\bibfield
  {journal} {\bibinfo  {journal} {Phys. Rev. A}\ }\textbf {\bibinfo {volume}
  {65}},\ \bibinfo {pages} {012509} (\bibinfo {year} {2001})}\BibitemShut
  {NoStop}%
\bibitem [{\citenamefont {Mitroy}\ and\ \citenamefont
  {Bromley}(2003{\natexlab{a}})}]{PhysRevA.67.034502}%
  \BibitemOpen
  \bibfield  {author} {\bibinfo {author} {\bibfnamefont {J.}~\bibnamefont
  {Mitroy}}\ and\ \bibinfo {author} {\bibfnamefont {M.~W.~J.}\ \bibnamefont
  {Bromley}},\ }\href {\doibase 10.1103/PhysRevA.67.034502} {\bibfield
  {journal} {\bibinfo  {journal} {Phys. Rev. A}\ }\textbf {\bibinfo {volume}
  {67}},\ \bibinfo {pages} {034502} (\bibinfo {year}
  {2003}{\natexlab{a}})}\BibitemShut {NoStop}%
\bibitem [{\citenamefont {Fabrikant}\ and\ \citenamefont
  {Gribakin}(2014)}]{PhysRevA.90.052717}%
  \BibitemOpen
  \bibfield  {author} {\bibinfo {author} {\bibfnamefont {I.~I.}\ \bibnamefont
  {Fabrikant}}\ and\ \bibinfo {author} {\bibfnamefont {G.~F.}\ \bibnamefont
  {Gribakin}},\ }\href {\doibase 10.1103/PhysRevA.90.052717} {\bibfield
  {journal} {\bibinfo  {journal} {Phys. Rev. A}\ }\textbf {\bibinfo {volume}
  {90}},\ \bibinfo {pages} {052717} (\bibinfo {year} {2014})}\BibitemShut
  {NoStop}%
\bibitem [{\citenamefont {Landau}\ and\ \citenamefont
  {Lifshitz}(1965)}]{landau}%
  \BibitemOpen
  \bibfield  {author} {\bibinfo {author} {\bibfnamefont {L.~D.}\ \bibnamefont
  {Landau}}\ and\ \bibinfo {author} {\bibfnamefont {E.~M.}\ \bibnamefont
  {Lifshitz}},\ }\href@noop {} {\emph {\bibinfo {title} {Quantum Mechanics:
  Non-Relativistic Theory}}},\ \bibinfo {edition} {2nd}\ ed.\ (\bibinfo
  {publisher} {Pergamon Press},\ \bibinfo {address} {Oxford},\ \bibinfo {year}
  {1965})\BibitemShut {NoStop}%
\bibitem [{\citenamefont {Ganas}(1972)}]{PhysRevA.5.1684}%
  \BibitemOpen
  \bibfield  {author} {\bibinfo {author} {\bibfnamefont {P.~S.}\ \bibnamefont
  {Ganas}},\ }\href {\doibase 10.1103/PhysRevA.5.1684} {\bibfield  {journal}
  {\bibinfo  {journal} {Phys. Rev. A}\ }\textbf {\bibinfo {volume} {5}},\
  \bibinfo {pages} {1684} (\bibinfo {year} {1972})}\BibitemShut {NoStop}%
\bibitem [{\citenamefont {Gribakin}\ and\ \citenamefont
  {Flambaum}(1993)}]{GF93}%
  \BibitemOpen
  \bibfield  {author} {\bibinfo {author} {\bibfnamefont {G.~F.}\ \bibnamefont
  {Gribakin}}\ and\ \bibinfo {author} {\bibfnamefont {V.~V.}\ \bibnamefont
  {Flambaum}},\ }\href {\doibase 10.1103/PhysRevA.48.546} {\bibfield  {journal}
  {\bibinfo  {journal} {Phys. Rev. A}\ }\textbf {\bibinfo {volume} {48}},\
  \bibinfo {pages} {546} (\bibinfo {year} {1993})}\BibitemShut {NoStop}%
\bibitem [{\citenamefont {Mitroy}\ and\ \citenamefont
  {Bromley}(2003{\natexlab{b}})}]{PhysRevA.68.035201}%
  \BibitemOpen
  \bibfield  {author} {\bibinfo {author} {\bibfnamefont {J.}~\bibnamefont
  {Mitroy}}\ and\ \bibinfo {author} {\bibfnamefont {M.~W.~J.}\ \bibnamefont
  {Bromley}},\ }\href {\doibase 10.1103/PhysRevA.68.035201} {\bibfield
  {journal} {\bibinfo  {journal} {Phys. Rev. A}\ }\textbf {\bibinfo {volume}
  {68}},\ \bibinfo {pages} {035201} (\bibinfo {year}
  {2003}{\natexlab{b}})}\BibitemShut {NoStop}%
\bibitem [{\citenamefont {Kumar}\ and\ \citenamefont
  {Meath}(1985{\natexlab{a}})}]{KM85}%
  \BibitemOpen
  \bibfield  {author} {\bibinfo {author} {\bibfnamefont {A.}~\bibnamefont
  {Kumar}}\ and\ \bibinfo {author} {\bibfnamefont {W.~J.}\ \bibnamefont
  {Meath}},\ }\href {\doibase 10.1080/00268978500103191} {\bibfield  {journal}
  {\bibinfo  {journal} {Mol. Phys.}\ }\textbf {\bibinfo {volume} {54}},\
  \bibinfo {pages} {823} (\bibinfo {year} {1985}{\natexlab{a}})}\BibitemShut
  {NoStop}%
\bibitem [{\citenamefont {London}(1937)}]{London}%
  \BibitemOpen
  \bibfield  {author} {\bibinfo {author} {\bibfnamefont {F.}~\bibnamefont
  {London}},\ }\href@noop {} {\bibfield  {journal} {\bibinfo  {journal} {Trans.
  Faraday Soc.}\ }\textbf {\bibinfo {volume} {33}},\ \bibinfo {pages} {8b}
  (\bibinfo {year} {1937})}\BibitemShut {NoStop}%
\bibitem [{\citenamefont {\mbox{Ya.} Amusia}\ \emph {et~al.}(1971)\citenamefont
  {\mbox{Ya.} Amusia}, \citenamefont {Cherepkov},\ and\ \citenamefont
  {Chernysheva}}]{amusiaJETP2}%
  \BibitemOpen
  \bibfield  {author} {\bibinfo {author} {\bibfnamefont {M.}~\bibnamefont
  {\mbox{Ya.} Amusia}}, \bibinfo {author} {\bibfnamefont {N.~A.}\ \bibnamefont
  {Cherepkov}}, \ and\ \bibinfo {author} {\bibfnamefont {L.~V.}\ \bibnamefont
  {Chernysheva}},\ }\href@noop {} {\bibfield  {journal} {\bibinfo  {journal}
  {Zh. Eksp. Teor. Fiz}\ }\textbf {\bibinfo {volume} {60}},\ \bibinfo {pages}
  {160} (\bibinfo {year} {1971})},\ \bibinfo {note} {[Sov. Phys. JETP, {\bf
  33}, 90 (1971)]}\BibitemShut {NoStop}%
\bibitem [{\citenamefont {\mbox{Ya}. Amusia}(1990)}]{atomicphotoeffect}%
  \BibitemOpen
  \bibfield  {author} {\bibinfo {author} {\bibfnamefont {M.}~\bibnamefont
  {\mbox{Ya}. Amusia}},\ }\href@noop {} {\emph {\bibinfo {title} {Atomic
  Photoeffect}}}\ (\bibinfo  {publisher} {Plenum, New York},\ \bibinfo {year}
  {1990})\BibitemShut {NoStop}%
\bibitem [{\citenamefont {Visser}\ \emph {et~al.}(1983)\citenamefont {Visser},
  \citenamefont {Wormer},\ and\ \citenamefont {Stam}}]{TDHF}%
  \BibitemOpen
  \bibfield  {author} {\bibinfo {author} {\bibfnamefont {F.}~\bibnamefont
  {Visser}}, \bibinfo {author} {\bibfnamefont {P.~E.~S.}\ \bibnamefont
  {Wormer}}, \ and\ \bibinfo {author} {\bibfnamefont {P.}~\bibnamefont
  {Stam}},\ }\href {\doibase 10.1063/1.445591} {\bibfield  {journal} {\bibinfo
  {journal} {J. Chem. Phys.}\ }\textbf {\bibinfo {volume} {79}},\ \bibinfo
  {pages} {4973} (\bibinfo {year} {1983})}\BibitemShut {NoStop}%
\bibitem [{\citenamefont {\mbox{Ya}. Amusia}\ \emph {et~al.}(1972)\citenamefont
  {\mbox{Ya}. Amusia}, \citenamefont {Cherepkov},\ and\ \citenamefont
  {Shapiro}}]{amusiaJETP}%
  \BibitemOpen
  \bibfield  {author} {\bibinfo {author} {\bibfnamefont {M.}~\bibnamefont
  {\mbox{Ya}. Amusia}}, \bibinfo {author} {\bibfnamefont {N.~A.}\ \bibnamefont
  {Cherepkov}}, \ and\ \bibinfo {author} {\bibfnamefont {S.}~\bibnamefont
  {Shapiro}},\ }\href@noop {} {\bibfield  {journal} {\bibinfo  {journal} {Zh.
  Eksp. Teor. Fiz.}\ }\textbf {\bibinfo {volume} {63}},\ \bibinfo {pages} {889}
  (\bibinfo {year} {1972})},\ \bibinfo {note} {[Sov. Phys. JETP, {\bf 36}, 468
  (1973)]}\BibitemShut {NoStop}%
\bibitem [{\citenamefont {Ludlow}(2003)}]{LudlowPhD}%
  \BibitemOpen
  \bibfield  {author} {\bibinfo {author} {\bibfnamefont {J.}~\bibnamefont
  {Ludlow}},\ }\href@noop {} {Ph.D. thesis},\ \bibinfo  {school} {Queen's
  University Belfast} (\bibinfo {year} {2003})\BibitemShut {NoStop}%
\bibitem [{\citenamefont {Berestetskii}\ \emph {et~al.}(1982)\citenamefont
  {Berestetskii}, \citenamefont {Lifshitz},\ and\ \citenamefont
  {Pitaevskii}}]{LandauQED}%
  \BibitemOpen
  \bibfield  {author} {\bibinfo {author} {\bibfnamefont {V.~B.}\ \bibnamefont
  {Berestetskii}}, \bibinfo {author} {\bibfnamefont {E.~M.}\ \bibnamefont
  {Lifshitz}}, \ and\ \bibinfo {author} {\bibfnamefont {L.~P.}\ \bibnamefont
  {Pitaevskii}},\ }\href@noop {} {\emph {\bibinfo {title} {Quantum
  Electrodynamics}}},\ \bibinfo {edition} {2nd}\ ed.\ (\bibinfo  {publisher}
  {Pergamon Press},\ \bibinfo {address} {Oxford},\ \bibinfo {year}
  {1982})\BibitemShut {NoStop}%
\bibitem [{\citenamefont {Gribakin}\ and\ \citenamefont
  {Ludlow}(2004)}]{PhysRevA.70.032720}%
  \BibitemOpen
  \bibfield  {author} {\bibinfo {author} {\bibfnamefont {G.~F.}\ \bibnamefont
  {Gribakin}}\ and\ \bibinfo {author} {\bibfnamefont {J.}~\bibnamefont
  {Ludlow}},\ }\href {\doibase 10.1103/PhysRevA.70.032720} {\bibfield
  {journal} {\bibinfo  {journal} {Phys. Rev. A}\ }\textbf {\bibinfo {volume}
  {70}},\ \bibinfo {pages} {032720} (\bibinfo {year} {2004})}\BibitemShut
  {NoStop}%
\bibitem [{\citenamefont {\mbox{Ya}. Amusia}\ and\ \citenamefont
  {Cherepkov}(1975)}]{csap}%
  \BibitemOpen
  \bibfield  {author} {\bibinfo {author} {\bibfnamefont {M.}~\bibnamefont
  {\mbox{Ya}. Amusia}}\ and\ \bibinfo {author} {\bibfnamefont {N.~A.}\
  \bibnamefont {Cherepkov}},\ }\href@noop {} {\bibfield  {journal} {\bibinfo
  {journal} {Case Studies in Atomic Physics}\ }\textbf {\bibinfo {volume}
  {5}},\ \bibinfo {pages} {47} (\bibinfo {year} {1975})}\BibitemShut {NoStop}%
\bibitem [{\citenamefont {Kolb}\ \emph {et~al.}(1982)\citenamefont {Kolb},
  \citenamefont {Johnson},\ and\ \citenamefont {Shorer}}]{PhysRevA.26.19}%
  \BibitemOpen
  \bibfield  {author} {\bibinfo {author} {\bibfnamefont {D.}~\bibnamefont
  {Kolb}}, \bibinfo {author} {\bibfnamefont {W.~R.}\ \bibnamefont {Johnson}}, \
  and\ \bibinfo {author} {\bibfnamefont {P.}~\bibnamefont {Shorer}},\ }\href
  {\doibase 10.1103/PhysRevA.26.19} {\bibfield  {journal} {\bibinfo  {journal}
  {Phys. Rev. A}\ }\textbf {\bibinfo {volume} {26}},\ \bibinfo {pages} {19}
  (\bibinfo {year} {1982})}\BibitemShut {NoStop}%
\bibitem [{\citenamefont {Schwerdtfeger}(2006)}]{onlinepolarizabilities}%
  \BibitemOpen
  \bibfield  {author} {\bibinfo {author} {\bibfnamefont {P.}~\bibnamefont
  {Schwerdtfeger}},\ }\enquote {\bibinfo {title} {Atomic static dipole
  polarizabilities},}\ in\ \href@noop {} {\emph {\bibinfo {booktitle}
  {Computational Aspects of Electric Polarizability Calculations: Atoms,
  Molecules and Clusters}}},\ \bibinfo {editor} {edited by\ \bibinfo {editor}
  {\bibfnamefont {G.}~\bibnamefont {Maroulis}}}\ (\bibinfo  {publisher} {IOS
  Press},\ \bibinfo {address} {Amsterdam},\ \bibinfo {year} {2006})\ pp.\
  \bibinfo {pages} {1--32},\ \bibinfo {note} {updated static dipole
  polarizabilities are available as a PDF file from the CTCP website at Massey
  University: http://ctcp.massey.ac.nz/dipole-polarizabilities.}\BibitemShut
  {Stop}%
\bibitem [{\citenamefont {Porsev}\ and\ \citenamefont
  {Derevianko}(2006)}]{porsevJETP}%
  \BibitemOpen
  \bibfield  {author} {\bibinfo {author} {\bibfnamefont {S.~G.}\ \bibnamefont
  {Porsev}}\ and\ \bibinfo {author} {\bibfnamefont {A.}~\bibnamefont
  {Derevianko}},\ }\href@noop {} {\bibfield  {journal} {\bibinfo  {journal}
  {Zh. Eksp. Teor. Fiz.}\ }\textbf {\bibinfo {volume} {129}},\ \bibinfo {pages}
  {227} (\bibinfo {year} {2006})},\ \bibinfo {note} {[Sov. Phys. JETP {\bf
  102}, 195 (2006)]}\BibitemShut {NoStop}%
\bibitem [{\citenamefont {McEachran}\ \emph {et~al.}(1979)\citenamefont
  {McEachran}, \citenamefont {Stauffer},\ and\ \citenamefont
  {Greita}}]{McESG79}%
  \BibitemOpen
  \bibfield  {author} {\bibinfo {author} {\bibfnamefont {R.~P.}\ \bibnamefont
  {McEachran}}, \bibinfo {author} {\bibfnamefont {A.~D.}\ \bibnamefont
  {Stauffer}}, \ and\ \bibinfo {author} {\bibfnamefont {S.}~\bibnamefont
  {Greita}},\ }\href {\doibase 10.1088/0022-3700/12/18/026} {\bibfield
  {journal} {\bibinfo  {journal} {J. Phys. B}\ }\textbf {\bibinfo {volume}
  {12}},\ \bibinfo {pages} {3119} (\bibinfo {year} {1979})}\BibitemShut
  {NoStop}%
\bibitem [{\citenamefont {Derevianko}\ \emph {et~al.}(2010)\citenamefont
  {Derevianko}, \citenamefont {Porsev},\ and\ \citenamefont
  {Babb}}]{Derevianko2010323}%
  \BibitemOpen
  \bibfield  {author} {\bibinfo {author} {\bibfnamefont {A.}~\bibnamefont
  {Derevianko}}, \bibinfo {author} {\bibfnamefont {S.~G.}\ \bibnamefont
  {Porsev}}, \ and\ \bibinfo {author} {\bibfnamefont {J.~F.}\ \bibnamefont
  {Babb}},\ }\href {\doibase http://dx.doi.org/10.1016/j.adt.2009.12.002}
  {\bibfield  {journal} {\bibinfo  {journal} {At. Data Nucl. Data Tables}\
  }\textbf {\bibinfo {volume} {96}},\ \bibinfo {pages} {323 } (\bibinfo {year}
  {2010})}\BibitemShut {NoStop}%
\bibitem [{\citenamefont {Martin}\ and\ \citenamefont
  {Fraser}(1980)}]{0022-3700-13-17-017}%
  \BibitemOpen
  \bibfield  {author} {\bibinfo {author} {\bibfnamefont {D.~W.}\ \bibnamefont
  {Martin}}\ and\ \bibinfo {author} {\bibfnamefont {P.~A.}\ \bibnamefont
  {Fraser}},\ }\href {http://stacks.iop.org/0022-3700/13/i=17/a=017} {\bibfield
   {journal} {\bibinfo  {journal} {J. Phys. B}\ }\textbf {\bibinfo {volume}
  {13}},\ \bibinfo {pages} {3383} (\bibinfo {year} {1980})}\BibitemShut
  {NoStop}%
\bibitem [{\citenamefont {Lide}(2009)}]{CRC}%
  \BibitemOpen
  \bibinfo {editor} {\bibfnamefont {D.~R.}\ \bibnamefont {Lide}},\ ed.,\
  \href@noop {} {\emph {\bibinfo {title} {{CRC} {H}andbook of {C}hemistry and
  {P}hysics}}},\ \bibinfo {edition} {89th}\ ed.\ (\bibinfo  {publisher} {CRC
  Press},\ \bibinfo {address} {Boca Raton, FL},\ \bibinfo {year}
  {2008--2009})\BibitemShut {NoStop}%
\bibitem [{\citenamefont {Porsev}\ and\ \citenamefont
  {Derevianko}(2002)}]{PhysRevA.65.020701}%
  \BibitemOpen
  \bibfield  {author} {\bibinfo {author} {\bibfnamefont {S.~G.}\ \bibnamefont
  {Porsev}}\ and\ \bibinfo {author} {\bibfnamefont {A.}~\bibnamefont
  {Derevianko}},\ }\href {\doibase 10.1103/PhysRevA.65.020701} {\bibfield
  {journal} {\bibinfo  {journal} {Phys. Rev. A}\ }\textbf {\bibinfo {volume}
  {65}},\ \bibinfo {pages} {020701} (\bibinfo {year} {2002})}\BibitemShut
  {NoStop}%
\bibitem [{\citenamefont {Kumar}\ and\ \citenamefont
  {Meath}(1985{\natexlab{b}})}]{kumar1985pseudo}%
  \BibitemOpen
  \bibfield  {author} {\bibinfo {author} {\bibfnamefont {A.}~\bibnamefont
  {Kumar}}\ and\ \bibinfo {author} {\bibfnamefont {W.~J.}\ \bibnamefont
  {Meath}},\ }\href@noop {} {\bibfield  {journal} {\bibinfo  {journal} {Mol.
  Phys.}\ }\textbf {\bibinfo {volume} {54}},\ \bibinfo {pages} {823} (\bibinfo
  {year} {1985}{\natexlab{b}})}\BibitemShut {NoStop}%
\bibitem [{\citenamefont {Mitroy}\ \emph {et~al.}(2002)\citenamefont {Mitroy},
  \citenamefont {Bromley},\ and\ \citenamefont {Ryzhikh}}]{MBR02}%
  \BibitemOpen
  \bibfield  {author} {\bibinfo {author} {\bibfnamefont {J.}~\bibnamefont
  {Mitroy}}, \bibinfo {author} {\bibfnamefont {M.~W.~J.}\ \bibnamefont
  {Bromley}}, \ and\ \bibinfo {author} {\bibfnamefont {G.~G.}\ \bibnamefont
  {Ryzhikh}},\ }\href {\doibase 10.1088/0953-4075/35/13/201} {\bibfield
  {journal} {\bibinfo  {journal} {J. Phys. B}\ }\textbf {\bibinfo {volume}
  {35}},\ \bibinfo {pages} {R81} (\bibinfo {year} {2002})}\BibitemShut
  {NoStop}%
\bibitem [{\citenamefont {Ludlow}\ and\ \citenamefont
  {Gribakin}(2010)}]{halogen}%
  \BibitemOpen
  \bibfield  {author} {\bibinfo {author} {\bibfnamefont {J.~A.}\ \bibnamefont
  {Ludlow}}\ and\ \bibinfo {author} {\bibfnamefont {G.~F.}\ \bibnamefont
  {Gribakin}},\ }\href@noop {} {\bibfield  {journal} {\bibinfo  {journal} {Int.
  Rev. At. Mol. Phys.}\ }\textbf {\bibinfo {volume} {1}},\ \bibinfo {pages}
  {73} (\bibinfo {year} {2010})},\ \bibinfo {note} {see also
  arXiv:physics/1002.3125v1}\BibitemShut {NoStop}%
\bibitem [{\citenamefont {Mitroy}\ and\ \citenamefont {Ryzhikh}(1999)}]{MR99}%
  \BibitemOpen
  \bibfield  {author} {\bibinfo {author} {\bibfnamefont {J.}~\bibnamefont
  {Mitroy}}\ and\ \bibinfo {author} {\bibfnamefont {G.}~\bibnamefont
  {Ryzhikh}},\ }\href {\doibase 10.1088/0953-4075/32/15/314} {\bibfield
  {journal} {\bibinfo  {journal} {J. Phys. B}\ }\textbf {\bibinfo {volume}
  {32}},\ \bibinfo {pages} {3839} (\bibinfo {year} {1999})}\BibitemShut
  {NoStop}%
\end{thebibliography}%

\end{document}